\documentclass[prl,twocolumn,preprintnumbers,amsmath,amssymb]{revtex4}
\usepackage{epsfig,amssymb,amsmath,psfrag}
\usepackage{hyperref}
\usepackage{pstricks,pst-node,graphicx}

\catcode`\@=11

\def \be  {\begin{equation}}
\def \ee  {\end{equation}}
\def \ba  {\begin{eqnarray}}
\def \ea  {\end{eqnarray}}
\def \baa {\begin{eqnarray*}}
\def \eaa {\end{eqnarray*}}
\def \bb  {\begin {thebibliography} }
\def \eb  {\end{thebibliography}}
\def \lab #1 {\label{#1}}
\newcommand{\ft}[2]{{\textstyle\frac{#1}{#2}}}

\def \matrix #1 {\left(\begin{array}{cc} #1 \end{array}\right)}

\newcommand{\as}{\ifmmode\alpha_{\rm s}\else{$\alpha_{\rm s}$}\fi}
\newcommand{\asbar}{\ifmmode\bar{\alpha}_{\rm s}\else{$\bar{\alpha}_{\rm s}$}\fi}

\font\cmss=cmss12 
\def\inbar{\,\vrule height1.5ex width.4pt depth0pt}
\def\IC{\relax\hbox{$\inbar\kern-.3em{\rm C}$}}
\def\IZ{\relax{\hbox{\cmss Z\kern-.4em Z}}}
\def\IR{{\hbox{{\rm I}\kern-.2em\hbox{\rm R}}}}

\def\IP{{\hbox{{\rm I}\kern-.2em\hbox{\rm P}}}}
\def\II{\hbox{{1}\kern-.25em\hbox{l}}}

\begin{document}
\title{An exact slope for AdS/CFT}
\author{B.~Basso,\footnote{bbasso@princeton.edu}}

\vspace{10mm}

\affiliation{Princeton Center for Theoretical Science, Jadwin Hall, Princeton University, Princeton, NJ 08544, USA}

\begin{abstract}
We present a conjecture for the small spin limit of the minimal scaling dimension of Wilson operators in the $\mathfrak{sl}(2)$ sector of the planar $\mathcal{N}=4$ Super-Yang-Mills theory. The expression is given in closed form as a function of the 't Hooft coupling and twist of the operator. The formula should stand as a prediction of the Asymptotic Bethe Ansatz equations for the spectrum of scaling dimensions and evidence is given at both weak and strong coupling that it should be exact. In particular, agreement is found with established one-loop spectroscopy of string energies at strong coupling.

\end{abstract}

\maketitle

{\it 1.~Introduction:} The AdS/CFT correspondence~\cite{Mal97,GKP98,W} reformulates the spectral problem of the planar maximally supersymmetric Yang-Mills (SYM) theory in terms of the spectrum of energies of a string propagating in the curved $\rm AdS_{5} \times S^{5}$ geometry. This statement alone has elucidated the fate of the scaling dimensions of the gauge theory at strong 't Hooft coupling $\lambda \equiv g^2_{\rm YM} N$. It predicted, in particular, the appearance of a large gap $\sim \lambda^{1/4} \gg 1$ separating scaling dimensions of protected and unprotected operators~\cite{GKP98}. It remained as a challenge for a long time to reproduce this gap starting from the gauge theory and to unravel the details of the weak-to-strong-coupling transition for the spectrum of excited states.

One of the first step in the program came with the seminal analysis of Berenstein, Maldacena, and Nastase (BMN)~\cite{BMN} where some of the difficulties of the gauge/string interpolation were bypassed by considering long enough operators in the gauge theory. This idea has triggered the development of the spin-chain approach where excited states are understood as made out of impurities/spin-waves propagating and scattering on top of a BPS background. As a remnant consequence of the properties of this background, perhaps, remarkable integrable structures have been uncovered -- see~\cite{Beisert} for a recent review -- that led eventually to a set of Asymptotic Bethe Ansatz (ABA) equations~\cite{BS05} which solve for the scattering egeinstates of the spin chain and compute the scaling dimensions of the associated conformal operators. Despite tremendous successes, the ABA equations failed for short enough operators or at strong enough coupling for which they ought to be supplied with the so-called wrapping corrections~\cite{AJK05,KLRSV}, which stand as finite-size corrections in the spin-chain/world-sheet picture. Later on, the ABA equations and the tower of wrapping corrections were successfully united and encoded into a set of Y system or TBA equations~\cite{GKV,AF}, which should embody all the dynamical features that dress the gauge and string theory into one another. Unfortunately, the Y-system/TBA equations are notoriously difficult to solve, and, to date, still await for their analytical solution -- see~\cite{D} for recent progress in this direction.

In this letter, we would like to point out that, apparently, some important information on the spectrum of scaling dimensions can be easily obtained, at any coupling, by working close enough to the BPS limit. Though very limited in extent and conjectural in nature, the result we shall present has the major advantage that it illustrates, in a relatively simple way, how the expected interpolation between the gauge and string theory takes place for short operators. The main purpose of this letter is to give credit to the expression that we shall propose.

To go deeper into the matter of this letter, we will be considering the so-called $\mathfrak{sl}(2)$ sector of Wilson operators $\sim \textrm{tr}\, D^{S}Z^J +$ mixing, where $D$ is a light-cone covariant derivative and $Z$ a complex scalar field. We shall restrict ourselves to the conformal operator carrying the minimal scaling dimension at given spin $S$ and twist $J$. The computation of this scaling dimension has been at the center of several studies, as it is one of the simplest, yet rich enough, probe of the AdS/CFT correspondence, in which it is mapped to a folded string rotating classically in $\rm AdS_{3} \times S^1$~\cite{GKP,FT02}. Our focus here is the leading correction to the minimal scaling dimension at small spin -- close to the BPS limit $S=0$ where the scaling dimension is exactly given by the twist. Our result is thus for the slope $\alpha_{J}$ of the minimal scaling dimension $\Delta_{J}(S)$ defined as
\ba\label{SmallSpin}
\Delta \equiv \Delta_{J}(S) = J + \alpha_{J} S + O(S^2)\, .
\ea
The slope $\alpha_{J}$ is function of the coupling $\lambda$ and twist $J$, normalized as $\alpha_{J} = 1 + O(\lambda)$ at weak coupling. It was previously introduced in~\cite{L} for the twist-two anomalous dimension where its expansion at weak coupling was constructed up to three loops. In this letter we would like to generalize this result to arbitrary twist and coupling.

{\it 2.~Exact expression:} The exact result for the slope is proposed to be
\ba\label{MainForm}
\alpha_{J} =  {\sqrt{\lambda}\over J}\, Y_{J}(\sqrt{\lambda}) \, ,
\ea
for arbitrary twist $J$ and at any coupling $\lambda$ in planar $\mathcal{N}=4$ SYM theory, where the function $Y_{J}(x)$ is defined as the ratio
\ba\label{Yj}
Y_{J}(x) \equiv {I'_{J}(x) \over I_{J}(x)}\, .
\ea
Here $I_{J}(x)$ stands for the $J^{\textrm{th}}$ modified Bessel's function, with scaling $I_{J}(x) \sim x^J/(2^JJ!)$ at small $x$, and $I'_{J}(x) \equiv d I_{J}(x)/dx$ for its derivative. The function $Y_{J}(x)$ has been introduced and studied in~\cite{Amos}, from which our notation is taken. Equivalent representation for the slope can be found as
\ba
\alpha_{J} = 1+{\sqrt{\lambda} \over J}{I_{J+1}(\sqrt{\lambda}) \over I_{J}(\sqrt{\lambda})} =  {\sqrt{\lambda} \over J}{I_{J-1}(\sqrt{\lambda}) \over I_{J}(\sqrt{\lambda})}-1\, ,
\ea
using known recurrence relations for the Bessel's functions.

Though the formula~(\ref{MainForm}) is easily handled in the various limits of interest, using the many well-known expansions for the Bessel's functions, it is often convenient to appeal to the non-linear first order differential equation solved by the function $Y_{J}(x)$ itself. This equation, derived in~\cite{Amos}, reads
\ba\label{DiffEq}
{dY_{J}(x) \over dx} = 1+{J^2 \over x^2} -{Y_{J}(x) \over x}-Y^{2}_{J}(x)\, .
\ea
Its general solution is of the form $Y_{J}(x) = d\log F_{J}(x)/dx$, where $F_{J}(x) = I_{J}(x) + \xi K_{J}(x)$, with $K_{J}(x)$ the second modified Bessel's function and $\xi$ an arbitrary parameter, independent on $x$. In this letter, only $\xi = 0$ is physical, but it is interesting, nevertheless, to consider what happens for $\xi \neq 0$. At weak coupling, for instance, we note that only $\xi=0$ is consistent with a gauge theory result: any other value of $\xi$ would affect the weak coupling limit yielding $\Delta = J-S + O(\lambda)$ in place of $\Delta = J+S + O(\lambda)$. We note, on the contrary, that the value of $\xi$ is almost irrelevant at strong coupling, if $\xi \neq \infty$. This is because this coefficient accompanies non-perturbative $\sim \exp{(-2\sqrt{\lambda})}$ corrections, lying far below a non-Borel summable tail of perturbative contributions.

{\it 3.~Origin of the expression:} The expression~(\ref{MainForm}) for twist two, i.e., $J=2$, originates from the solution to the ABA equations. The expression~(\ref{MainForm}) at arbitrary twist $J$ is proposed as its natural extension. To be more precise, the expression for twist two was obtained by solving the long-range Baxter equation~\cite{B06} in the $\mathfrak{sl}(2)$ sector of operators under study. The latter equation, whose dynamical content (and thus range of validity) is equivalent to the ABA equations, offers a better setting for the analytical continuation at small spin. The construction of this solution will be produced in a future publication~\cite{B}. We expect this proof can be extended to encompass higher twist as well.

We should stress, however, that even if the formula~(\ref{MainForm}) for $J=2$ shall be given as a prediction of the Baxter equation in~\cite{B}, it remains of interest to compare its predictions with other constructions, like the ones obtained by directly continuing to small spin the expressions found at arbitrary integer spin using the ABA equations. The reason is that the uniqueness of the physical solution to the Baxter equation is not entirely settled away from integer spin values, leaving some  plausible doubt about the validity of~(\ref{MainForm}). We expect, however, that the solution that shall be given in~\cite{B}, due to its elegant simplicity, is the physical one. 

The ABA equations are not complete, due to missing wrapping corrections, and one might expect that the formula~(\ref{MainForm}), even if
confirmed as a definite prediction of the ABA equations, could not be correct. There are few cases, mostly related to the large spin limit of scaling dimensions, where wrapping effects appear to be irrelevant. The most well-known example is the cusp anomalous dimension~\cite{KR85,P80}, for which an exact equation, controlling its dependence on the coupling, was produced starting from ABA~\cite{BES}. This observable relates to a Wilson loop expectation value and is thus intrinsically tied to Feynman diagrams with the topology of the disk, for which the notion of wrapping does no apply. In the present case, however, the absence of wrapping corrections is not apparently under control. Nevertheless, as surprising as it is, all the explicit results for wrapping corrections known to date~\cite{BJL,LRV,BFTZ}, i.e., for twist two and minimal twist three scaling dimension, reveal that wrapping corrections are suppressed at small spin, starting at $\sim S^2$. This observation leads us to our assumption that the expression~(\ref{MainForm}) is exact in the coupling.

We should mention at this point that the explicit results for wrapping corrections, referred to before, only capture the so-called first L\"uscher contribution. The scaling $\sim S^2$ at small spin of this contribution could well be an accidental property, valid at this level only. To put our assumption on firm ground one should extend consideration to the TBA or Y-system equations. It is unfortunately not known to us how to perform the (necessary) analytical continuation to small spin of these equations. It would be of course of tremendous interest to develop such an approach, not only for testing the absence of wrapping corrections in~(\ref{MainForm}), but also for more ambitious challenges as the ones raised by the comparison with the Balitsky-Fadin-Kuraev-Lipatov (BFKL) Pomeron -- see~\cite{KLRSV} and references therein -- which also entail working away from natural integer spins.

By staying within the scope of the present study, we shall be able to perform some tests of the `exactness' of~(\ref{MainForm}), beyond the first L\"uscher correction, by going at strong coupling and matching with established one-loop spectroscopy of strings. But first, let us proceed with the consideration of the weak coupling limit.

{\it 4.~Weak coupling analysis:} The weak coupling expansion of~(\ref{MainForm}) reads
\ba\label{WCE}
\alpha_{J} = 1+{\lambda \over 2J(1+J)}-{\lambda^2 \over 8J(1+J)^2(2+J)} + O(\lambda^3)\, ,
\ea
and follows straightforwardly from the representation in terms of the Bessel's function. More generally, the weak coupling expansion is convergent, with a radius of convergency determined by the position of the first (non-trivial) zero of the Bessel's function $I_{J}(\sqrt{\lambda})$, which is reached at some imaginary value of $\sqrt{\lambda}$.

For twist two, the expression~(\ref{WCE}) is found to be in perfect agreement with the result reported in~\cite{L}. In this reference, the slope was derived up to three loops from explicit result for the twist-two anomalous dimension. We have performed a similar analysis for the minimal twist-three anomalous dimension up to three loops, using expression obtained in~\cite{KLRSV}, and observed agreement with~(\ref{WCE}). This type of comparison is unfortunately restricted to these two particular cases for which closed expression for any integer spin can be constructed explicitely. An interesting exception is the large twist limit where the BMN result should be recovered. It reads
\ba\label{largeJ}
\alpha_{J} = 1 + {\lambda \over 2J^2} + O(1/J^3)\, ,
\ea
and it immediately follows from the large $J$ asymptotics $Y_{J}(\sqrt{\lambda}) = J/\sqrt{\lambda} + \sqrt{\lambda}/(2J) + O(1/J^2)$. Interestingly, the first subleading correction to~(\ref{largeJ}), which is equal to $-\lambda/(2J^3)$, matches exactly with the spin-chain result of~\cite{BTZ} (see also~\cite{fnote} and Appendix~D of~\cite{MTT}).

The weak coupling expansion derived from~(\ref{MainForm}) has a remarkably simple transcendental pattern. Indeed, when performed in terms of the coupling $g = \sqrt{\lambda}/(4\pi)$, which is conventional in the spin-chain description, it only involves coefficients which are suitable (loop-order dependent) powers of $\pi^2$, up to overall rational numbers. This may appear quite surprising given that an important ingredient of the ABA equations, namely the dressing phase~\cite{BES,BHL}, is obviously parameterized, at weak coupling, in terms of infinitely many zeta values with odd arguments. The latters are not reducible to powers of $\pi^2$ and usually show up in any weak coupling expression at a high enough order in pertubation theory. The point here is that the derivation of~(\ref{MainForm}) from the long-range Baxter equation reveals that the dressing phase contribution scales away at small spin. It only affects the scaling dimension at order $\sim S^2$, as can be easily seen at twist two by looking at the dressing phase contribution to the four-loop anomalous dimension~\cite{KLRSV}. The absence of dressing phase contribution could well be responsible for a large part to the simplicity of the result~(\ref{MainForm}).

{\it 5.~Intermediate coupling analysis:} There is no simple way to study analytically the slope at intermediate coupling. The Bessel's functions are however well-tabulated functions and therefore their numerical estimate is easily obtained. A sample of the numerical plots of the slope $\alpha_{J}$ as a function of the coupling $\sqrt{\lambda}$ is depicted in Fig.~1, for low values of the twist $J$. It shows that the transition from weak to strong coupling is rather smooth. The smoothness of the transition could well explain the relatively good agreement with the Pad\'e approximation used in~\cite{L} to extrapolate and resum the perturbative expansion of the slope of the twist two anomalous dimension.

We further note the similarity with the numerical construction of the cusp anomalous dimension~\cite{Benna06}.
\begin{figure}[t]
\includegraphics[width=70mm]{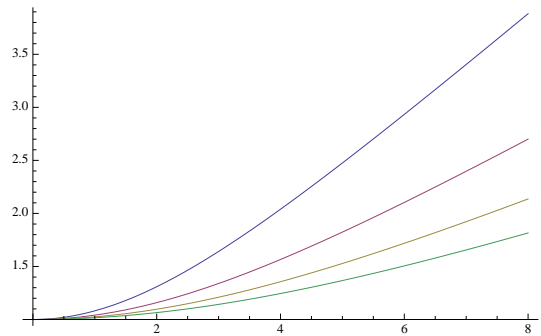}
\caption{Plot of the slope $\alpha_{J}$ as a function of the coupling $\sqrt{\lambda}$ for $J = 2, \ldots , 5$. The upper (blue) line stands for $J=2$. The slope decreases monotonically as a function of the twist $J$ at fixed coupling}
\end{figure}


{\it 6.~Strong coupling analysis:} The analysis of the slope at strong coupling is also very straightforward. Starting from the general expression~(\ref{MainForm}) one easily derives that
\ba\label{ASC}
\alpha_{J} = {\sqrt{\lambda} \over J} - {1 \over 2J} + O(1/\sqrt{\lambda})\, .
\ea
In distinction with weak coupling, the strong coupling expansion is asymptotic and non-Borel summable, i.e., its Borel resummation is ambiguous. Perhaps the easiest way to understand it is to look at the differential equation~(\ref{MainForm}). Solving this equation as an expansion in $1/x$ -- we recall that $x = \sqrt{\lambda}$ -- leaves no freedom up to a branch choice which is solved by imposing that $\alpha_{J} $ should be positive. However, this equation is first order in $x$ and thus as a zero-mode ambiguity, associated to the parameter $\xi$ introduced before. The independence of the strong coupling expansion on this parameter indicates that the strong coupling expansion cannot uniquely fix the solution or does it only up to $\exp{(-2\sqrt{\lambda})}$ corrections -- and (integer) powers therefore due to the non-linearity of the equation. These corrections are the ones associated to the ambiguity in the Borel resummation of the expansion~(\ref{ASC}). Note that, in this regard, the case $\xi = \infty$ is exceptional, since then the perturbative series becomes Borel summable. Its strong coupling expansion is obtained from~(\ref{ASC}) upon the substitution $\sqrt{\lambda} \rightarrow -\sqrt{\lambda}$. It corresponds then to an unphysical branch at strong coupling with a negative asymptotics for the slope.

Unfortunately, there is a serious drawback concerning the value of the result~(\ref{ASC}) from a string theory perspective. It comes along with the question of the range of validity of the small spin expansion at strong coupling. The large behavior of the slope at strong coupling indicates that the radius of convergency gets asymptotically small at strong coupling. This is not a direct consequence of our analysis, strictly speaking. A proper treatment would require consideration of contributions $\sim S^n$, with $n > 1$, which should definitely receive wrapping corrections. The discussion, here and below, is actually adapted from the analysis of the BFKL Pomeron at strong coupling~\cite{BPSI} --see also~\cite{C} for a nice account-- which we `minimally' extend to twist $J$ greater than $2$. The conclusion is that the small spin expansion has indeed a small radius of convergency $\sim 1/\sqrt{\lambda}$ at strong coupling. It prevents us from drawing any valuable conclusion for string energies with spin $\sim 1$ -- the small spin expansion is apparently not a good starting point at strong coupling or need to be resummed first. The situation is not so hopeless actually. The analysis below uses extra information on (expected) analytical properties of the scaling dimension to improve the expansion and make a new series of test of~(\ref{MainForm}).

To condense all our extra hypothesis in a closed form we shall start instead with the identity
\ba\label{Marginality}
\Delta^2 = J^2 + f_{J}(S)\, ,
\ea
where $f_{J}(S)$ has a Taylor expansion in $S$,
\ba\label{Texp}
f_{J}(S) = \beta_{J}S + O(S^2)\, .
\ea
The mapping of our previous result with~(\ref{Marginality}) leads to
\ba\label{beta-alpha}
\beta_{J} = 2J\alpha_{J} = 2\sqrt{\lambda}Y_{J}(\sqrt{\lambda}).
\ea
Our main assumption now is that the coefficients in the expansion~(\ref{Texp}) are more and more suppressed with the coupling, at strong coupling -- see formula~(\ref{TLexp}) below for illustration. This is, of course, in structural agreement with the semiclassical matching/resummation, and, thus, with the one-loop result obtained in~\cite{GSSV,RT,VM}, which has passed serious test~\cite{GSSV}. It is implying that the radius of convergency of~(\ref{Texp}) is as large as $\sim \sqrt{\lambda}$ at strong coupling, which should be associated to the presence of singularities in the complex spin plane at a distance $\sim \sqrt{\lambda}$ from the origin $S=0$. The latter singularities should control the flat-to-curved-space transition of the string energy, which we assume here is well captured by the semiclassical analysis~\cite{GKP, FT02} -- then providing a radius of convergency of order $\sim 1$ in the semiclassical spin variable $\mathcal{S} \equiv S/\sqrt{\lambda}$. Under this battery of assumptions, it follows that at order $n$ in the strong coupling expansion, performed at finite spin, one is allowed to truncate the expansion~(\ref{Texp}) to the first $n$ terms. We have then, to the leading order,
\ba\label{LO}
\Delta^2 = \beta_{J}S +O(1) = 2\sqrt{\lambda}S+O(1)\, ,
\ea
for arbitrary finite twist and spin. We thus reproduced the expected gap in the spectrum of scaling dimension at strong coupling, in perfect agreement with~\cite{GKP98,AFS,GSSV,RT,VM}. This illustrates the way we are going to peform test of~(\ref{beta-alpha}) at higher loops. Prior to present these results, let us come back to the original issue.

We now understand better that the closest singularity in the complex spin plane of the scaling dimension flows to the origin $S=0$ at strong coupling. Its position indeed is determined by the solution to the equation~(\ref{Marginality}) with $\Delta = 0$, which we denote by $S_{0}$,
\be
S_{0} = -{J^2 \over \beta_{J}} + O(1/\lambda^{3/2}) = -{J^2 \over 2\sqrt{\lambda}} + O(1/\lambda)\, .
\ee
For twist two, $J=2$, one recovers of course the strong coupling result~\cite{BPSI} for the Pomeron intercept $j_{0} = 2+S_{0} = 2-2/\sqrt{\lambda} + O(1/\lambda)$. By going to the parameterization~(\ref{Marginality}) we thus resolved the singularity $\Delta \sim \sqrt{S-S_{0}}$ and, by assuming that all other singularities are large $\sim \sqrt{\lambda}$, we were able to relate the asymptotics at strong coupling to the leading term in the Taylor expansion~(\ref{Texp}). A synthesis of our assumption concerning the complex spin plane for the minimal scaling dimension is depicted in Fig.~2. At weak coupling the situation is entirely different. In this case we expect $S_{0} = -J+1+O(\lambda)$, based on a linear interpolation of small and large $J$ result, and the small spin expansion has a radius of convergency $\sim J-1$. We expect furthermore that the extra singularities will collide with the one at $S_{0}$ such that there is no real improvement in going to the parametrization~(\ref{Marginality}) -- as far as the small spin analysis is concerned.

\begin{figure}[t]
\includegraphics[width=70mm]{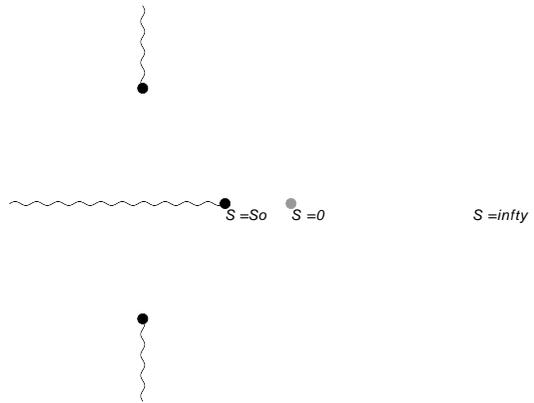}
\caption{Sketch of the (expected) complex spin plane for minimal scaling dimension. The central grey dot marks the origin -- it is not the locus of a singularity at finite coupling. The closest singularity lies on the negative $S$-axis at a position $S_{0} < 0$. Extra singularities are depicted in the upper and lower half plane -- they are expected to lie at a distance $\sim \sqrt{\lambda}$ from the origin. Presumably all singularities collide at $S = -J +1$ at weak coupling.}
\end{figure}

We will now present a set of consistency checks of the formula~(\ref{beta-alpha}), and, simultaneously, of  the lines of thoughts that led us to it, by means of a direct confrontation with several stringy results.

The first test concerns the higher-loop corrections to~(\ref{LO}). Using the large argument asymptotics of the Bessel's functions, or, better, solving the equation~(\ref{DiffEq}) at large $x$, we immediately find that
\ba\label{FJL}
\beta_{J} = 2\sqrt{\lambda}-1 + {J^2-\ft{1}{4} \over \sqrt{\lambda}} + {J^2-\ft{1}{4} \over \lambda} + O(1/\lambda^{3/2})\, .
\ea
The first two terms in the left-hand side of~(\ref{FJL}) are in perfect agreement with the one-loop stringy result of~\cite{GSSV,RT,VM}. We believe this matching represents a rather non-trivial test that the expression~(\ref{beta-alpha}) is exact in the coupling. If confirmed, the expansion~(\ref{FJL}) will stand for higher-loop prediction for the string theory. Looking at it more carefully, one would notice that the $n$-loop expansion coefficient in~(\ref{FJL}) is a polynomial in $J$ of degree $n$ -- with symmetry $J\rightarrow -J$, apparently. This feature makes the expansion compatible with a semiclassical resummation, which computes the scaling dimension at fixed semiclassical boost $\mathcal{J} \equiv J/\sqrt{\lambda}$, at strong coupling. This brings us to our second test.

By including more terms in~(\ref{FJL}) one can easily convinces him/herself that the result resums into
\ba\label{SCL}
\begin{aligned}
\beta_{J} = \, \, &2\sqrt{\lambda}\sqrt{1+\mathcal{J}^2}-{1\over 1+\mathcal{J}^2}\\
&\, \, \, + {\mathcal{J}^2-\ft{1}{4} \over \sqrt{\lambda}\, (1+\mathcal{J}^2)^{5/2}}+O(1/\lambda)\, .
\end{aligned}
\ea
The leading term is in perfect agreement with the classical string energy of a folded string, with small classical spin $\mathcal{S} \equiv S/\sqrt{\lambda}$, but arbitrary boost $\mathcal{J}$, which is known to be given by~\cite{FT02}
\ba
\left({\Delta \over \sqrt{\lambda}}\right)^2_{\textrm{classical}} = \mathcal{J}^2 + 2\sqrt{1+\mathcal{J}^2}\mathcal{S} + O(\mathcal{S}^2)\, .
\ea
The linear dependence on the spin $\mathcal{S}$ is therefore correctly reproduced by our formula at the classical level. The next terms in the right-hand side of~(\ref{SCL}) are higher-loop semiclassical predictions. Remarkably enough, the first of these corrections can be shown to be in perfect agreement with the prediction of the one-loop algebraic curve~\cite{Kolya}, which confirms the absence of wrapping corrections for the slope, due to non-trivial cancellations. Finally, the large $\mathcal{J}$ asymptotics of~(\ref{SCL}) is also found to match with stringy results~\cite{PTT,MTT}.

Note that we do not observe an order of limit issue, here. One could rederive the result~(\ref{SCL}) by solving the equation~(\ref{DiffEq}) directly in the relevant regime $\sqrt{\lambda} \gg 1$ with $\mathcal{J}$ fixed. The sought solution would be found as $Y_{J}(\sqrt{\lambda}) = \sqrt{1+\mathcal{J}^2} + \ldots$ from which the same expansion as in~(\ref{SCL}) would follow. It is interesting to note that in this approach the leading classical asymptotics directly originates from the algebraic equation
\ba\label{AlgEq}
0  = 1+\mathcal{J}^2-Y^2_{J}(\sqrt{\lambda})\, ,
\ea
to which the complete equation~(\ref{DiffEq}) reduces in this regime. That the equation becomes algebraic classically is directly related to our previous remark on the fact that the zero-mode ambiguity is associated with non-perturbative corrections at strong coupling. The perturbative tail of corrections is uniquely defined by the equation~(\ref{DiffEq}), with no further input on 'the initial condition' than the choice of the positive branch in~(\ref{AlgEq}), and thus the tail itself cannot define uniquely the physical solution away from strong coupling.

{\it 7. Conclusion and outlook:} In this letter, we have presented a formula for the slope of the minimal scaling dimension at small spin, for arbitrary coupling and twist, in planar $\mathcal{N}=4$ SYM theory. We gave support to the conjecture that it stands as an exact result for the AdS/CFT correpondence. The evidence included certain non-trivial matching with string theory results at strong coupling, together with the observation of the absence of wrapping contributions at weak coupling for twist two and three. We recall here that the formula~(\ref{MainForm}) applies for the $\mathfrak{sl}(2)$ sector only. A natural question to ask next is whether similar expressions exist in the compact $\mathfrak{su}(2)$ and fermionic $\mathfrak{su}(1|1)$ sectors.
 
We believe that the formula~(\ref{MainForm}) also sheds some light on the nature of the strong coupling expansion for finite spin and twist. Based on our expression, it seems reasonable to think that the strong coupling expansion will be asymptotic only. This seems even more reasonable if we assume that the expansion is indeed resummable order by order semiclassically. Then the typical scale of the $n$-loop semiclassical result would be of the order of the $n$-loop contribution to $\beta_{J}$ in~(\ref{beta-alpha}), which we know exhibits a factorial growth at large $n$. It is much less clear to us whether the divergent series will maintain its non-Borel summability. If it is case, it would be interesting to understand the physical origin of the non-perturbative contributions, as was done for the cusp anomalous dimension~\cite{AM07,BKK07}.

We would like also to mention that the formula~(\ref{beta-alpha}) can be used to partially fix the two-loop scaling dimension at strong coupling, for any finite spin and twist. This can be done by including the relevant subleading terms in the small spin expansion~(\ref{Texp}). It yields
\be\label{TLexp}
\begin{aligned}
\Delta^2 = J^2 &+ \left(2\sqrt{\lambda}- 1 + {J^2-\ft{1}{4} \over \sqrt{\lambda}}\right)S \\
&+ \left({3\over 2}-{b\over\sqrt{\lambda}}\right)S^2 - {3\over 8\sqrt{\lambda}}S^3 + O(1/\lambda) \, .
\end{aligned}
\ee
We already used in this expression the matching with the classical string energy~\cite{GKP,FT02} to fix the leading coefficients in front of $S^2$ and $S^3$. The quantity $b$, however, remains undetermined -- it is one loop semiclassically speaking and it shoud be independent on $J$. A direct comparison with the numerical estimate of the $J=S=2$ scaling dimension in the Konishi multiplet~\cite{GKV-k,Frolov} gives $b=3$ as a plausible integer candidate. It would be interesting to fix the value of $b$ by pushing a bit further the analysis of~\cite{GSSV}.

Finally, an intriguing question is whether the equation~(\ref{DiffEq}) that controls the slope can be derived directly in the world-sheet theory.

{\it Acknowledgements:} I would like to thank G.~Korchemsky and A.~Belitsky for interesting correspondence and careful reading of the manuscript. It is also a pleasure to acknowledge to A.~Zhiboedov for his friendship interest in this project.

\end{document}